\documentclass[dvips]{article}
\usepackage{icrctc07}

\title{Multi-wavelength Observations of PG\,1553+113 with HESS}
\shorttitle{Multi-wavelength Observations of PG\,1553+113}
\authors{
W.\,Benbow$^{1}$, 
C.\,Boisson$^{2}$, 
R.\,B\"uhler$^{1}$,
and H.\,Sol$^{2}$
for the HESS Collaboration
}
\shortauthors{W.\,Benbow et al.}
\afiliations{$^1$ Max-Planck-Institut f\"ur Kernphysik, Heidelberg, Germany\\ 
$^2$ LUTH, UMR 8102 du CNRS, Observatoire de Paris, Section de Meudon, France\\
}
\email{Wystan.Benbow@mpi-hd.mpg.de}

\abstract{
Very high energy (VHE; $>$100 GeV) $\gamma$-ray
observations of PG\,1553+113 were made with the 
High Energy Stereoscopic System (HESS) in 2005 and 2006.
A strong signal, $\sim$10 standard deviations, 
is detected by HESS during the 2 years of
observations (24.8 hours live time). 
The time-averaged energy spectrum, measured
between 225 GeV to $\sim$1.3 TeV,
is characterized
by a very steep power law (photon index of 
$\Gamma = 4.5\pm0.3_{\rm stat}\pm0.1_{\rm syst}$).
The integral flux above 300 GeV is $\sim$3.4\% of the Crab Nebula
flux and shows no evidence for any variations, on any time scale.
H+K (1.45$-$2.45$\mu$m) spectroscopy of PG\,1553+113 
was performed in March 2006 with SINFONI, an integral field 
spectrometer of the ESO Very Large Telescope (VLT) in Chile.
The redshift of PG\,1553+113 is still unknown, as
no absorption or emission lines were found.}

\begin{document}
\maketitle

\section{Introduction}

Evidence for VHE ($>$100 GeV) $\gamma$-ray emission from 
the high-frequency-peaked BL\,Lac object PG\,1553+113 was 
first reported by the HESS collaboration \cite{HESS_discovery} 
based on observations made in 2005. This detection was later
confirmed \cite{MAGIC_1553} with MAGIC observations 
in 2005 and 2006.   The measured VHE spectra are unusually soft 
(photon index $\Gamma$=4.0$\pm$0.6 and $\Gamma$=4.2$\pm$0.3 for the
HESS and MAGIC experiments, respectively) but the errors are large,
clearly requiring improved measurements before detailed 
interpretation of the complete SED is possible.
Further complicating any SED interpretation is the 
absorption of VHE photons \cite{EBL_effect3,EBL_effect2}
by pair-production on the Extragalactic Background Light (EBL).
This absorption, which is energy dependent and increases strongly 
with redshift, distorts the VHE energy spectra observed from 
distant objects. For a given redshift, the
effects of the EBL on the observed spectrum can be 
reasonably accounted for during SED modeling.
Unfortunately, the redshift of PG\,1553+113 
is unknown, despite many attempts to measure it
(see, e.g., \cite{Carangelo_03,no_lines}).

In 2005 and 2006, a total of 30.3 hours of HESS observations 
were taken on PG\,1553+113. The 2005 HESS observations are exactly the 
same as reported in \cite{HESS_discovery}.
The good-quality exposure is 24.8 hours live time.
The data are processed using the
standard HESS calibration \cite{calib_paper}
and analysis tools \cite{std_analysis}.
{\it Soft cuts} \cite{HESS_discovery}
are applied to select candidate $\gamma$-ray events,
resulting in an average post-analysis energy 
threshold of 300 GeV at the mean zenith angle of the observations, $37^{\circ}$. 

\section{HESS Results}

   \begin{figure}[t]
   \centering
      \includegraphics[width=0.45\textwidth]{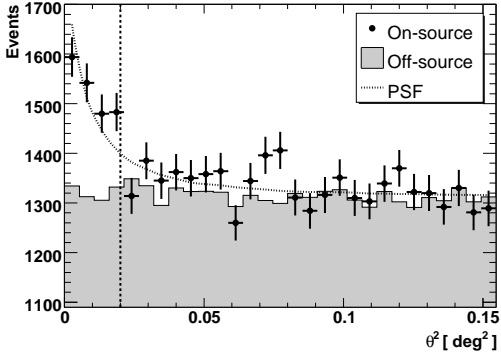}
      \caption{The distribution of $\theta^2$ for on-source 
        events (points) and
        normalized off-source events (shaded) from observations
        of PG\,1553+113.  The dashed curve represents
	the $\theta^2$ distribution expected for
	a point source of VHE $\gamma$-rays at $40^{\circ}$
	zenith angle with a photon index $\Gamma=4.5$.
	The vertical line represents the cut 
        on $\theta^2$ applied to the data.}
         \label{thtsq_plot}
   \end{figure}

  \begin{table}
      \caption{Shown are the excess, the significance of the excess, and 
        the integral flux above 300 GeV, from HESS observations
	of PG\,1553+113. The flux units are $10^{-12}$ cm$^{-2}$\,s$^{-1}$.
	The systematic error on the flux is 20\% and is not shown.}
         \label{results}
        \centering
         \begin{tabular}{c c c c c}
	   \\
            \hline\hline
            \noalign{\smallskip}
            Epoch & Time & Excess & Sig & I($>$300 GeV)\\
            & [h] & &  [$\sigma$] & [f.u.]\\
            \noalign{\smallskip}
            \hline
            \noalign{\smallskip}
            2005 & 7.6  & 249 & 6.0 & $5.44\pm1.23$\\
            2006 & 17.2 & 536 & 8.3 & $4.22\pm0.72$\\
            \noalign{\smallskip}
            \hline
            \noalign{\smallskip}
            Total & 24.8 & 785 & 10.2 & $4.56\pm0.62$\\             
            \noalign{\smallskip}
            \hline
       \end{tabular}
   \end{table}

A significant VHE $\gamma$-ray signal is detected 
in each year of HESS data taking.  The total observed
excess is 785 events, corresponding
to a statistical significance of 10.2 standard deviations ($\sigma$).
Table~\ref{results} shows the results of the HESS 
observations of PG\,1553+113. 
Figure~\ref{thtsq_plot} shows the on-source and normalized off-source
distributions of the square of the angular difference between
the reconstructed shower position and
the source position ($\theta^{2}$) for all observations. 
The background is approximately flat in $\theta^{2}$ as expected, and
there is a clear point-like excess of on-source events 
at small values of $\theta^{2}$, 
corresponding to the observed signal.
The peak of a two-dimensional Gaussian fit to a sky map 
of the observed excess is coincident with the position 
of PG\,1553+113.

\begin{figure}
   \centering
      \includegraphics[width=0.45\textwidth]{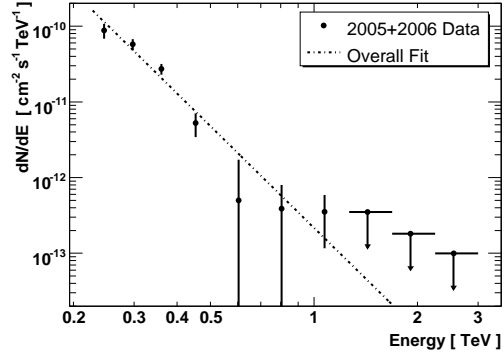}
      \caption{The overall VHE energy spectrum observed from PG\,1553+113. 
	The dashed line represents the best $\chi^2$ fit of a power law to
        the observed data.  The upper limits are at the 99\% confidence
	level \cite{UL_tech}. Only the statistical errors are shown.}
         \label{spectrum_plot}
   \end{figure}

The photon spectrum for the entire data set is shown 
in Figure~\ref{spectrum_plot}. These data are well fit
($\chi^2$ of 8.4 for 5 degrees of freedom) by
a power-law function, d$N$/d$E \sim E^{-\Gamma}$,
with a photon index  
$\Gamma=4.5\pm0.3_{\rm stat}\pm0.1_{\rm syst}$.
Fits of either a power law with an exponential cut-off
or a broken power law do not yield significantly
better $\chi^2$ values.  

\begin{figure}
   \centering
      \includegraphics[width=0.45\textwidth]{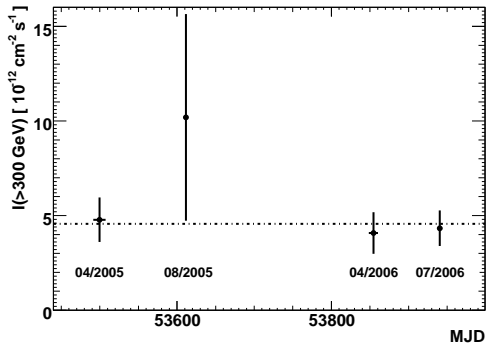}
      \caption{The integral flux, I($>$300 GeV), measured by HESS
from PG\,1553+113 during each dark period of observations.
The horizontal line represents the average flux for all 
the HESS observations. Only the statistical errors are shown.}
         \label{monthly_plot}
   \end{figure}

The observed integral flux above 300 GeV for the entire data set is
I($>$300 GeV) = $(4.6\pm0.6_{\rm stat}\pm0.9_{\rm syst}) \times 10^{-12}$ 
cm$^{-2}$\,s$^{-1}$.  This corresponds to $\sim$3.4\% of I($>$300 GeV)
determined from the HESS Crab Nebula spectrum \cite{HESS_crab}.
The integral flux, I($>$300 GeV), is shown in Table~\ref{results}
for each year of observations.   Figure \ref{monthly_plot} shows the flux 
measured for each dark period. There are no indications for flux variability 
on any time scale within the HESS data.
The data previously published \cite{HESS_discovery}
for HESS observations of PG\,1553+113 in 2005 were
not corrected for long-term changes in the optical sensitivity
of the instrument.  Relative to a virgin telescope, the total
optical throughput was decreased by 29\% in 2005 and 33\%
in 2006.  Correcting \cite{HESS_crab} for this decrease, 
using efficiencies determined from simulated and
observed muon events, increases the flux measured from the object.
The effect of this correction is larger for soft spectrum sources
than it is for hard spectrum sources. Due to the correction, 
the flux measured in 2005 is three times
higher than previously published.

On July 24 and 25, 2006, PG\,1553+113 
was observed by the Suzaku X-ray satellite (\cite{Suzaku_info},
Suzaku Observation Log: http://www.astro.isas.ac.jp/suzaku/index.html.en).
On these two dates HESS observed PG\,1553+113 for 3.1 hours live time,
resulting in a marginally significant excess of 101 events (3.9$\sigma$). 
The average flux measured on these two nights is I($>$300 GeV) = 
$(5.8\pm1.7_{\rm stat}\pm1.2_{\rm syst}) \times 10^{-12}$ 
cm$^{-2}$\,s$^{-1}$.  

\section{SINFONI Near-IR Spectroscopy}

The determination of the redshift of an AGN is
generally based upon the detection of emission or
absorption lines in its spectrum.
In an attempt to detect absorption features from the host galaxy or
emission lines from the AGN,  H+K (1.50--2.40$\mu$m) spectroscopy 
of PG\,1553+113 was performed with SINFONI, an integral field 
spectrometer mounted at Yepun, Unit Telescope 4 of the 
ESO Very Large Telescope in Chile. The source was observed
on March 9, 2006 and March 15, 2006. The resulting images are spatially
unresolved and no underlying host galaxy is detected.
 
   \begin{figure}
   \centering
      \includegraphics[width=0.3\textwidth,angle=270]{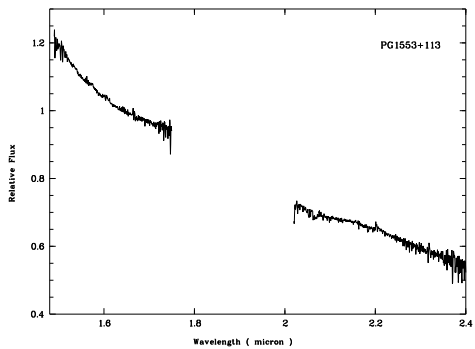}
      \caption{The measured H+K-band spectrum of PG\,1553+113
in relative flux units. The gap is due to the highly 
reduced atmospheric transmission between H and K bands. }
         \label{IR_spectrum}
   \end{figure}

The H+K-band spectrum of PG\,1553+113 measured here is shown in
Figure~\ref{IR_spectrum}. The signal-to-noise reach is $\sim$250 in
the H-band and $\sim$70 in the K-band.  The observed near-IR
spectrum is featureless apart from some residuals from the
atmospheric corrections. Thus in neither the broad-band images nor
in the spectrum are the influences of the gas of a host galaxy or
the AGN detected, even though PG\,1553+113 is bright in the IR.  As
a result, a redshift determination from these observations is not
possible.

\section{Discussion}

\begin{figure}
   \centering
      \includegraphics[width=0.45\textwidth]{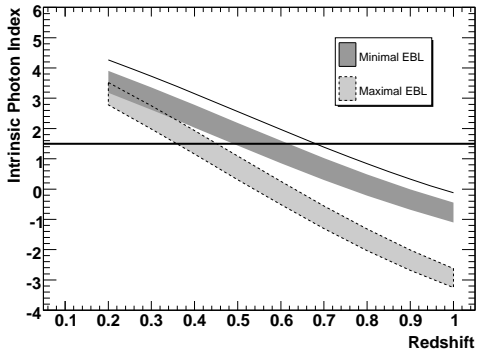}
      \caption{
The photon index $\Gamma_{\rm int}$ determined from a power-law fit to 
the intrinsic spectrum of PG 1553+113 (i.e. the H.E.S.S. data de-absorbed with an EBL model) 
for a range of redshifts. The contours reflect the 1$\sigma$ statistical uncertainty 
of the fits. The upper curve is the sum of $\Gamma_{\rm int}$ for the ¡ÈMinimal'' 
model and twice its statistical error.}
         \label{int_spec}
   \end{figure}

As the redshift of PG 1553+113 is likely $z>0.2$,
the observed VHE spectrum should be strongly affected by VHE $\gamma$-ray 
absorption on the EBL. If the redshift were known the spectrum intrinsic
to the source could be reconstructed assuming a model of the EBL density. 
However, the EBL SED is not well-determined. Using a {\it Maximal} 
EBL model at the level of the upper limits from \cite{Nature_EBL} 
or a {\it Minimal} model near the EBL lower limits 
from galaxy counts \cite{galaxy_counts} can yield a significantly different intrinsic spectrum.  
Figure~\ref{int_spec} shows the intrinsic spectrum versus redshift
for both the {\it Maximal} and {\it Minimal} EBL parameterizations.
Here, scaled models of \cite{Primack_EBL} are used, exactly as described 
in \cite{HESS_discovery}.  The redshift of the AGN can be limited 
using assumptions for the intrinsic spectrum. 
Assuming the intrinsic photon index is not harder than $\Gamma_{\rm int}=1.5$, 
a limit of $z<0.69$ is thus determined from the {\it Minimal} EBL model.

\section{Conclusion}

With a $\sim$25 h data set, $\sim$3 times larger than previously
published \cite{HESS_discovery}, the HESS signal from 
PG\,1553+113 is now highly 
significant ($\sim$10$\sigma$). Thus the evidence 
for VHE emission from PG\,1553+113 
previously reported is now clearly verified.
However, the flux observed in 2005 is now $\sim$3 times higher 
than initially reported due to an improved calibration 
of the absolute energy scale of HESS, and agrees well with
the flux measured in 2005 by MAGIC \cite{MAGIC_1553}. 
The statistical error on the VHE photon index is still rather large
($\sim$0.3), primarily due to the extreme softness of the 
observed spectrum ($\Gamma=4.5$).  
The total HESS exposure on PG\,1553+113 is $\sim$25 hours.  Barring
a flaring episode, not yet seen in two years of observations,
a considerably larger total exposure ($\sim$100 hours) would be
required to significantly improve the spectral measurement.
However, the VHE flux from other AGN is known to vary dramatically
and even a factor of a few would reduce the observation
requirement considerably. Should such a VHE flare occur, 
not only will the error on the measured VHE spectrum be 
smaller, but the measured photon index may 
also be harder (see, e.g., \cite{VHE_hardening}).  Both effects would 
dramatically improve the redshift constraints and correspondingly
the accuracy of the source modeling. Therefore, the VHE flux
from PG\,1553$+$113 will continue to be monitored by HESS.
In addition, the soft VHE spectrum makes it an ideal target for
the lower-threshold HESS Phase-II \cite{HESSII} which should
make its first observations in 2009.

\section{Acknowledgements}
The support of the Namibian authorities and of the University of Namibia
in facilitating the construction and operation of H.E.S.S. is gratefully
acknowledged, as is the support by the German Ministry for Education and
Research (BMBF), the Max Planck Society, the French Ministry for Research,
the CNRS-IN2P3 and the Astroparticle Interdisciplinary Programme of the
CNRS, the U.K. Science and Technology Facilities Council (STFC),
the IPNP of the Charles University, the Polish Ministry of Science and 
Higher Education, the South African Department of
Science and Technology and National Research Foundation, and by the
University of Namibia. We appreciate the excellent work of the technical
support staff in Berlin, Durham, Hamburg, Heidelberg, Palaiseau, Paris,
Saclay, and in Namibia in the construction and operation of the
equipment. Based on ESO-VLT SINFONI program 276.B-5036 observations.


\end{document}